\documentclass[11pt]{article}
\usepackage{cospar}
\usepackage{url}

\usepackage{graphicx}
\usepackage[figuresright]{rotating}

\title{BAX: A dedicated X-Rays galaxy clusters Database}

\author{ R. Sadat \address{LAOMP, CNRS, Observatoire Midi-Pyr\'en\'ees,
 14 Av. Edouard Belin, 31 400, Toulouse, France,}
 A. Blanchard$^{1}$, C. Mendiboure$^{1}$, and J.-P. Kneib$^{1}$}

\begin{document}

\maketitle

\begin{abstract}
We present BAX, Base de Donn\'ees amas de galaxies X (http://webast.ast.obs-mip.fr/bax), a project aiming at building a comprehensive database dedicated
 to X-rays clusters of galaxies allowing detailed information retrieval. BAX 
provides the user with {\it basic data} published in the literature on X-rays 
clusters of galaxies as well as with information concerning the physical 
properties in the X-rays domain or at other wavelengths. BAX allows individual 
studies on selected clusters as well as building up homogenous samples, from  known X-rays 
clusters for which selection criteria are chosen through web interfaces. 
We expect BAX to become a useful tool for astronomy community in order to 
optimize the cluster science return using data from both ground based 
facilities like MEGACAM (CFHT), VIRMOS (VLT) and space missions like XMM, 
Chandra and Planck.  \\

\end{abstract}

\section*{INTRODUCTION}

X-rays observations of clusters of galaxies have matured rapidly in the last ten
 years with the advent of good quality imaging and spectroscopy on board 
several space observatories (Ginga, Tenma, ROSAT, ASCA, XMM, Chandra) . All 
these observations resulted in large amount of X-rays data for thousands groups
 and clusters of galaxies published in an exponentially increasing number of 
papers.\\ 
As clusters of galaxies are extended objects that can be observed at several 
wavelengths, their study result in increasing number of excellent scientific X-rays data and publications a dedicated database is therefore strongly needed in order to optimize their scientific exploitation.\\ 
By now, there exist no X-rays clusters dedicated database. Although some 
information (catalogs, tables, bibliographic references...) can be retrieved 
from existing databases such as CDS, VIZIER or NED, more specific information 
on X-rays clusters properties can not be obtained.  
Such a database is strongly needed and should be particularly appreciated by 
the astronomical community in order to extract rapidly X-rays data on galaxy clusters in order to analyze and understand the global 
properties of these objects and to study the correlations between fundamental 
physical quantities or in the preparation of observational run on a specific set of clusters. BAX is also an online research database designed to 
support scientists, space missions and ground based observatories in the 
observations planning, data analysis and publication of research on galaxy clusters. 

\section*{GENERAL DESCRIPTION OF BAX CAPABILITIES}

BAX contains :\\
- cluster's coordinates ($\alpha$, $\delta$) and the cluster redshift $z$ \\
- selected observational measurements: the cluster X-rays flux ($Fx$), X-rays
luminosity ($Lx$) and X-rays temperature ($Tx$)\\
- a set of bibliographical references . \\
Depending on the menu query, the user can search for a given cluster name or a set of clusters that meet a choosen selection criteria, this criteria can be either on cluster position or on the observational measurements or on some bibliographical keywords. 
For each published measurement, BAX gives the bibliographic references with a World Wide Web (WWW) interface link to the {\it Abstract Service of the Astrophysics Data System} (ADS) paper (Kurtz et al. 2000).

\section*{HOW DOES BAX OPERATE ?}

The major tables in BAX database are the following:\\ 
--   published bibliographic references of interest, described by their {\it bibcode} as used by ADS (Kurtz et al. 2000),\\ 
--  primary names of  clusters (internal) with their acronyms as obtained from
NED, \\
--   measurements of the basic data ($\alpha$, 
$\delta$, $z$, $Fx$, $Lx$, $Tx$) and the corresponding keywords describing 
the contents of the bibliographic references on each cluster. 
Administration of the BAX database can be performed 
either by reading an Ascii 
file with a standard format,  or through a World Wide Web interface.
In the first step BAX integrates for each cluster the 
acronyms, the coordinates and the redshift which are obtained by running NED scripts (the ``Byname'' 
and ``Bypositions'' batch jobs). In the second step BAX integrates the 
measurements of the cluster flux, $Fx$, the cluster luminosity $Lx$ and the cluster temperatrure 
$Tx$) as well as the relevant wavelength band of observation for $Fx$ and $Lx$.  $Lx$ is automatically computed from the flux using the redshift existing in BAX, as well as, eventually, the flux in a fixed wavelength 
band, the ROSAT [0.1-2.4keV] band in order to provide homogenous ``canonical values'' ($Fx$, $Lx$, 
$Tx$) for each cluster. These quantities are provided to simplify the 
use of BAX, and should be treated as so. If used, this is under the 
responsibility of the user. Finally, 
 keywords associated to the couple (cluster name, bibliographic reference) 
are integrated.  

\subsection*{Query Functions Through the WEB Interface}

\begin{center}
\begin{minipage}{145mm}
\includegraphics[width=145mm]{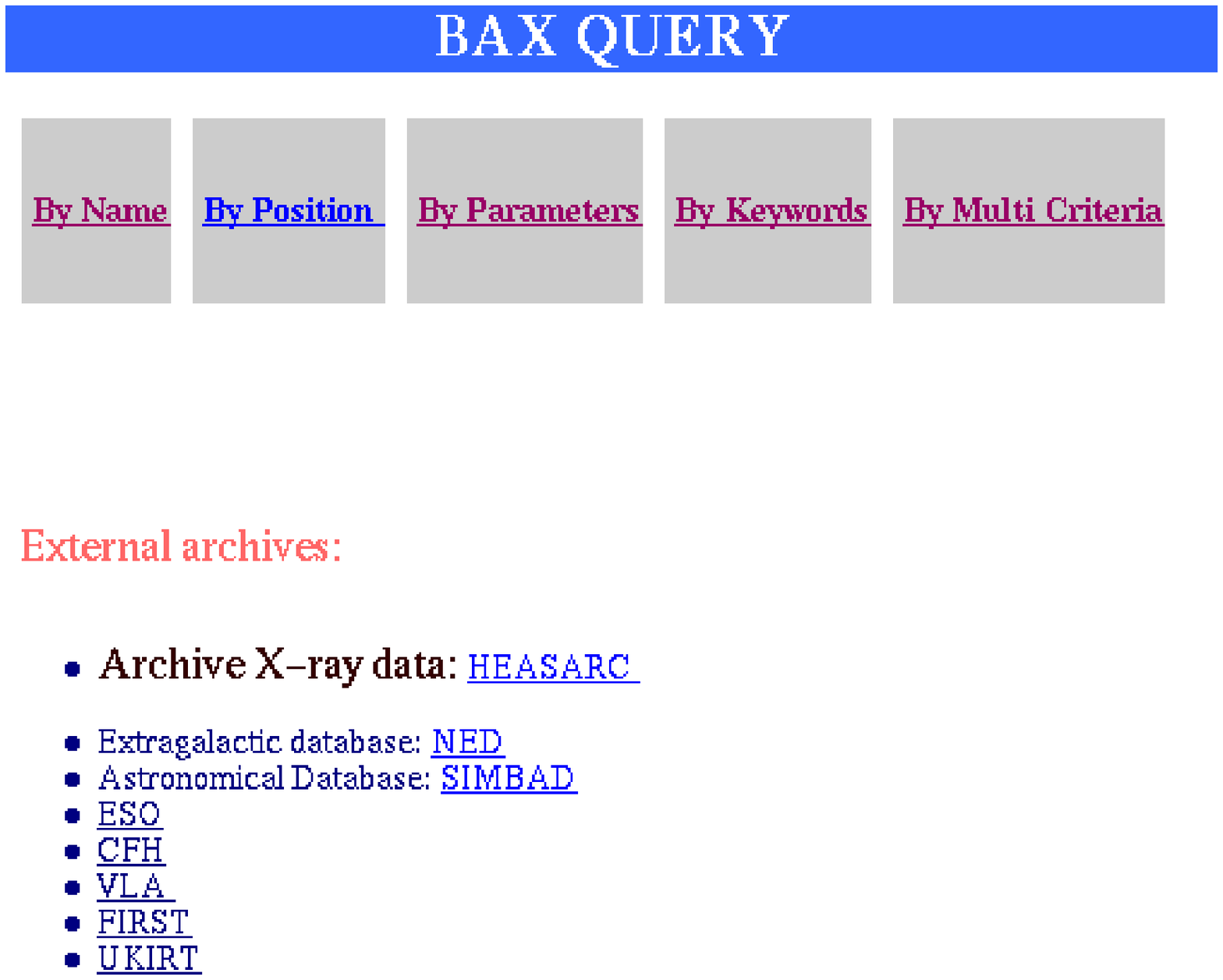}

{\sf Fig. 1. Front page of the BAX query interface. The various query menus 
proposed by BAX. }

\end{minipage}
\end{center}

\vspace*{10mm}%

The BAX database is accessible through the World Wide Web interface  at the
following address:\\ 
{ \centerline{\sf http://webast.ast.obs-mip.fr/bax}}

 The user starts 
his query by selecting the query menu through the main page. 
By now five query menus exist: ``By name query'', ``By positions query'', 
`` By parameters query'', ``By keywords query'' and a more complex menu 
which combines all the menus called ``By multi-criteria'' menu designed for 
search by name, positions, physical quantities measurements and keywords 
respectively. BAX will then retrieve information included in the data base
 as well as their references, with their associated links, corresponding
to a given 
cluster or a list of clusters that respond to one or several criteria. For 
each of the query menu (except for the ``keywords menu'') the primary 
outputs are: 
the cluster positions, the cluster redshift $z$ and one set of measurements 
of the three observational types : $Fx$ and  $Lx$ in a ROSAT reference band,
 and $Tx$) qualified as {\it 
canonical measurements} which are selected among the most precise and/or 
the most recent ones. The user can go on asking for more information to get 
access to all published measurements through the {\it all measurements} query 
and/or access to an ensemble of bibliographic references selected on keywords 
criteria concerning this cluster through the {\it bibliographic query}. 
In each mode, BAX provide only a limited number of answers, while the comprehensive answer is provided as an ascii file.  
Figure 1 shows the BAX interface main menu, with the five query 
menus.
One of the main motivation of building BAX database is to allow to retrieve  a 
sample of clusters selected on the basis of one or several criteria in order 
to perform statistical analysis or for the purpose of a systematic study. 
For example, BAX can provide very easily the sample of clusters 
above a given limited flux with available temperature measurements (provided the
informations have been published): this can be obtained by a simple query of BAX using the joint 
constraints on the flux $Fx$ and the temperature $Tx$. 
A more complex query can also be performed. For example the user can search for a set of ABELL clusters in a given area of the sky 
for which temperature measurements, temperature map and weak lensing data exist with the relevant references.
\subsection*{CURRENT STATUS }

BAX is already fully accessible. By now it contains more than a thousand clusters with basic data and among which 264 are with temperature measurements.

\section*{FUTURE IMPROVEMENTS}

In its present form, BAX is restricted to a database. However, 
the project includes a following period intend  to develop 
additional services in order to provide useful knowledges on X-rays clusters, either directly
or, when existing, as links through the WEB, forming a portal on the scientific subject.  
In this perspective, in addition to provide basic informations on galaxy clusters, BAX will be 
able to offer new 
functionalities to improve its capabilities:
\noindent -- enhancement in the BAX search by adding the batch mode.\\ 
--  tools for studying X-rays clusters: XMM data analysis using SASsoftware, tools to visualize the query outputs (histograms, 
charts...)\\
-- External links: query and extraction of relevant information in an automated fashion. Access to X-rays archives \\

\section*{CONCLUSIONS}

BAX provides with published data measurements of the basic properties of  clusters including pointers 
to related bibliographical references. The ultimate goal of BAX is to provide not only an interface to 
query its database but also to serve as a portal for clusters of galaxies research community.
We expect BAX to become an essential tool for the astronomical community especially in the view of 
preparing Planck mission which will detect up to 10,000 individual galaxy clusters up to $z \approx 1$.  

\section*{ACKNOWLEDGMENTS}

BAX has been supported by the Centre National de la Recherche Spatiale (CNES), the french Programme National de Cosmologie (PNC) and the Observatoire de Midi-Pyr\'en\'ees (OMP). 
We thank the NED team for their help and more specifically Joe Mazzarella for 
support on the NED-client scripts and B. Madore.  
We are very grateful to C. Faure and Y. Loiseau who have contributed as graduate students to the early phase of BAX development.
Finally, we want to thank M. Arnaud, D. Egret, J.P. Kneib, G. Mathez and J. Bartlett who greatly helped to define the contents and functions of BAX.

\end{document}